\documentclass[namedreferences]{kluwer}

\usepackage{epsfig}
\usepackage{times}
\usepackage{mathptm}

\input epsf
\def\eps@scaling{.95}
\def\epsscale#1{\gdef\eps@scaling{#1}}
\def\plotone#1{\centering \leavevmode
    \epsfxsize=\eps@scaling\columnwidth \epsfbox{#1}}

\newcommand{\um}{\,\hbox{$\mu$m}}
\def\deg{\nobreak{$^\circ$}}
\def\arcsec{\nobreak{$''$}}
\newcommand{\mv}{\hbox{m$_{\rm V}$}}
\def\spose#1{\hbox to 0pt{#1\hss}}
\def\simlt{\mathrel{\spose{\lower 3pt\hbox{$\mathchar"218$}}
     \raise 2.0pt\hbox{$\mathchar"13C$}}}
\def\simgt{\mathrel{\spose{\lower 3pt\hbox{$\mathchar"218$}}
     \raise 2.0pt\hbox{$\mathchar"13E$}}}

\begin{document}

\begin{article}

%
%

\begin{opening}

\title{The ALFA Laser and Analysis Tools}

\author{S. \surname{Rabien}\email{srabien@mpe.mpg.de}}
\institute{Max-Planck-Institut f\"ur extraterrestrische Physik,
Garching, Germany}
\author{T. \surname{Ott}}
\author{W. \surname{Hackenberg}}
\author{A. \surname{Eckart}}
\author{R. \surname{Davies}}
\institute{Max-Planck-Institut f\"ur extraterrestrische Physik,
Garching, Germany}
\author{M. \surname{Kasper}}
\institute{Max-Planck-Institut f\"ur Astronomie, Heidelberg, Germany}
\author{A. \surname{Quirrenbach}}
\institute{University of California, San Diego, USA}

%
\begin{abstract}
The optimal performance of adaptive optics systems can only be maintained
if the wavefront reference is bright and compact.
Experience has shown that both of these important criteria are remarkably
difficult to achieve with laser guide stars, and 
this contribution gives an account of the methods by which ALFA
attempts to reach them.
First, the production of a high quality, high power laser beam is
described.
However, this quality is unavoidably compromised along the path to the
launch telescope.
In order to rectify this, a new set of diagnostic tools which monitor
the quality of the out-going beam has been installed, and these are
also described.
Lastly, we outline a number of possible modifications on which we are
working.
If successful, these may allow a substantial improvement in the beam
quality.
\end{abstract}



\end{opening}

%
%

\section{Introduction} 
\label{sect:intro}

\begin{figure}
\plotone{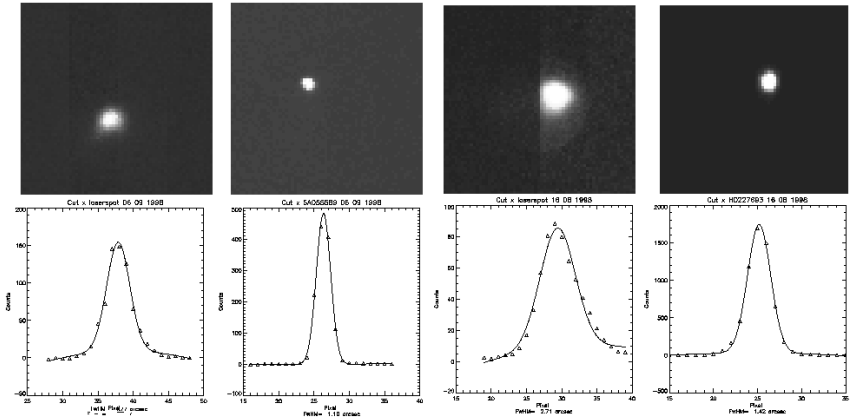}
\caption{Laser spot and natural guide stars imaged on the wave front
sensor under different conditions: 
A) LGS in good transparency 
B) NGS with the same conditions as (A) and \mv=9.4 
C) LGS while observing the laser through clouds 
D) NGS with the same conditions as (C) with \mv=9.0.}
\label{fig:spots}
\end{figure}

The aim of the ALFA laser guide star (LGS) is the generation of an
artificial reference star for the adaptive optics (AO), in order to
increase the sky coverage of the system. 
For best results the Shack-Hartmann sensor of the AO system needs a
sufficiently bright compact source to obtain centroid measurements with
a good signal-to-noise ratio. 
So the main objective is: 
how to get the brightest and smallest LGS image on the wavefront sensor. 
The heart of the ALFA laser facility is an argon ion pumped dye jet
laser, which is installed in the coud\'e lab at the Calar Alto 3.5-m
telescope. 
The laser emits light at the wavelength of the sodium D$_2$ line with an
output power up to 4\,W, with extremely high beam quality.
In order to create an artificial guide star in the mesospheric sodium
layer, the laser beam is transported from the laser laboratory along
the coud\'e path of the 3.5-m to a Galilean beam expander near the main
mirror of the telescope, where it is finally launched (see Ott et al.,
this issue). 
Along this path, a distance of approximately 50\,m, the laser is 
subjected to the disturbing influence of the turbulent air inside the
dome and other distortion sources, so that the beam quality is
significantly worse than when it leaves the coud\'e lab.
The aberrations can affect the brightness and compactness of the laser
guide star, as shown in Figure~\ref{fig:spots}, which are so critical
to the performance of the adaptive optics (see Davies et al., this
issue).
A series of diagnostic tools have therefore been installed directly
beneath the launch telescope to allow monitoring of the
quality of the projected beam.

\section{The Laser}
\label{sec:laser}

\begin{figure}
\epsscale{1.1}
\plotone{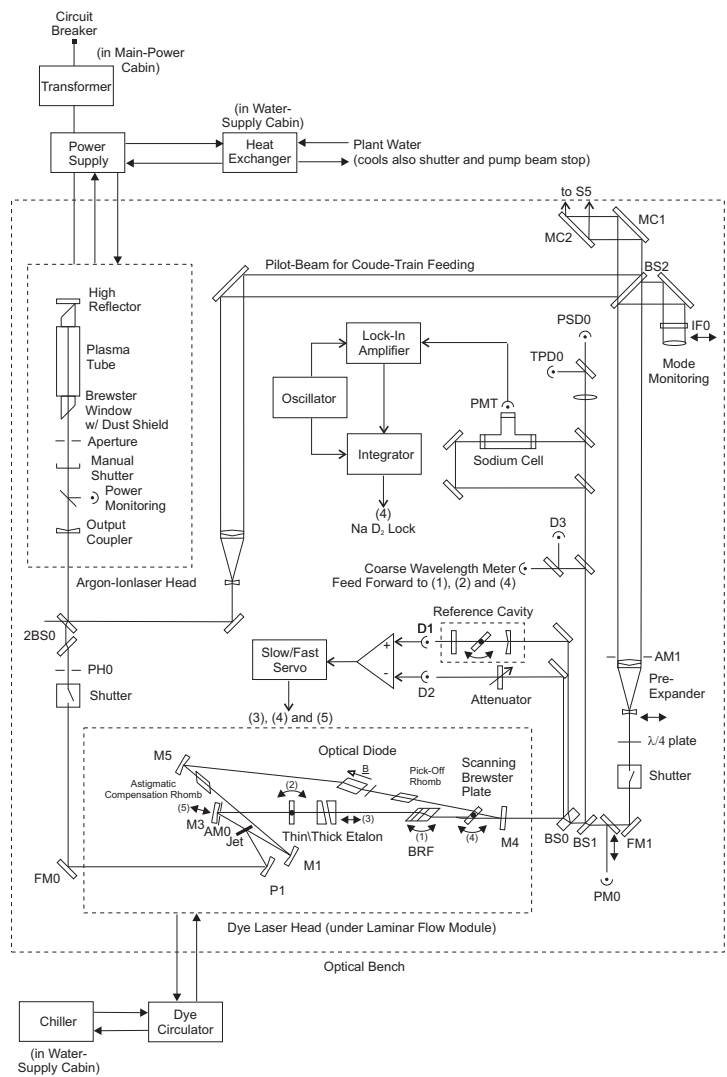}
\caption{Layout of the 1.2\,m$\times$2.4\,m laser optical bench in the
  coud\'e lab. On the
  left is the Ar$^+$ laser. Some of the output is used for alignment,
  but most pumps the dye laser. A small part of the dye laser output is
  used for frequency tuning and the rest passes through a shutter,
  quarter-wave plate, and pre-expander before being fed along the
  coud\'e train.}
\label{fig:laserbench}
\end{figure}

The main component of the ALFA laser system is a Coherent 899 ring dye
laser, pumped by a Coherent INNOVA 200 Argon ion laser.
These lasers, together with optics for beam forming and diagnostics,
are mounted on a floated optical bench (Figure~\ref{fig:laserbench}) in
a separate room on the coud\'e floor, and shielded with removeable covers. 
Above the dye laser a laminar flow box is installed to prevent the
sensitive resonator components from contamination by, for example, dust
in the air. 
The laser room is equipped with an air conditioning system to stabilize
the air temperature to 20$\pm$1\deg C independent of the
environmental conditions.

The lasing medium in the ring dye laser is Rhodamine 6G (Rh6G). 
For exiting Rh6G, intense blue or green light is necessary. 
Today only continuous wave ion lasers are able to produce high enough
power in that wavelength regime in a good single-spatial mode. 
The quality of the pump laser (power stability, mode structure) is
important because the dye laser output is influenced by the input of
the pump laser as well as the flow characteristics of the dye jet.
The multiline output of the ALFA pump laser consists of eight lines
from 457.9--514.5\,nm which lase simultaneously so that, although their
relative line strengths are not ideally matched to the absorption band
of Rh6G, the power output is maximised.

The pump laser provides 27\,W of multi-line output, about 2\,W of which
is fed directly into the coud\'e train to be
used as a pilot beam for alignment, and the rest pumps the dye ring
laser.
This is sufficient for a dye laser output of typically 3.8\,W, with
single spatial and temporal mode quality.
Higher powers (up to 5.5\,W) can be achieved but at the cost of
considerably higher maintenance overheads \cite{qui97a}.

\subsection{Argon-Ion pump laser}

The cavity of the pump laser consists of a low pressure plasma tube
with ionized argon gas as the active lasing medium positioned between
one flat high reflector and one curved output coupler. 
The adjustable high reflector is sealed onto the plasma tube. 
Between the front Brewster window and the output coupler there are an
intracavity aperture, a manual shutter and a beamsplitter for power
monitoring. 
The intracavity aperture is adjusted for the largest aperture diameter
that provides the highest output power (for a given tube current) with
a Gaussian TEM$_{\rm 00}$ mode. 

The plasma tube of the pump laser is cooled in a closed loop water
cooling system (flow rate 25\,l\,min$^{-1}$). 
The heat exchanger, as well as the chiller for the dye solution are
installed in the water supply cabin which is adjacent to the laser lab
to prevent interference with the running system. 
The water tubes are shielded by separate pipes which can withstand a
tube leakage, and are mounted under the floor of the laser cabin.

The microprocessor controlled power supply of the pump laser provides
the DC current for the tube and its axial electro-magnet, and
is equipped with a circuit breaker and a 480/380\,V transformer. 
It is located in different room and monitors safety interlocks (covers,
water flow) and diagnoses system faults. 
Communication is either via a serial link or with a remote
control module.

During normal operation of the system, the pump laser status, its tube
current, tube voltage, and output power are monitored continuously
using a serial line.
Changing any of these parameters is only possible using the remote
control module which is located next to the pump laser in the laser
laboratory, so as to prevent damage from excessive tinkering.




\subsection{Ring Dye Laser}

The basic design of any continuous wave dye laser is constrained by the
photophysics and chemistry of the  dissolved organic dye molecules,
which represent the active laser medium. 
The absorption and emission spectra of dissolved dye molecules are
broadband features, a result of the fact that in a dye solution the
closely spaced rotational-vibrational levels are heavily collision
broadened such that they overlap. 
The absorption band at lower frequencies is nearly the mirror image of
the emission band, and the exact peak position of these bands depend on
the solvent and dye concentration.

The broadening of the gain of organic dyes is both a blessing and a
curse. 
It is a blessing in the sense that in the case of homogenous broadening
all of the gain medium can contribute power to the oscillating laser
mode, and because the broad emission spectrum provides the laser's
tunability. 
It is a curse because the broad spectral width means that the lifetime
of the excited-state is short and hence intense pump powers are
required in order to achieve sufficient population inversion for the
laser oscillation. 

The major elements of the ring dye laser are: 
\begin{itemize}
\item
The resonator which is a figure-8-shaped ring, formed with 3
spherical mirrors and one flat output coupler with 10\% transmission
\item
A curved mirror which focuses the pump light into the free
flowing dye jet stream, which is the active medium of the dye laser 
\item
The dye circulator which cools, filters and pumps the dye solution
\end{itemize}

Because of the heat produced in the dye by the focused pump light which
is not re-radiated as fluorescence and also because of triplet-state
trapping, it is necessary to have the dye molecules traverse the pump
spot very rapidly.
The excited dye molecules can decay into both singlet- and triplet-states.
But due to the much longer lifetime of the triplet-states, these act as
a trap for the excited dye molecules, which are then no more available
for the lasing process. 
Therefore the flow velocity at the output of the jet-forming nozzle of
the dye laser is such that the dye molecules stay only in the pump spot
for a time similar to the triplet-lifetime.
Additionally, triplet-state trapping can further be quenched chemically
by adding some amount of cyclo-octatetraene (COT) to the dye solution.

The dye solution is kept at a constant temperature with a chiller
installed in the water supply cabin. 
The materials coming into contact with the dye in the circulation
system (stainless steel, teflon) are chosen such that they do not
influence the lifetime of Rh6G, which is therefore  subject to
degradation only due to pump-laser heating and laser-induced
photochemistry.

The transmitted fraction of the pump light (several Watts) falls onto a
water-cooled beam-stop in the dye laser head. 
The pump laser shutter is automatically closed in case of a failure of
the dye circulation system.

A number of modifications (dye solvent, flow
velocity, temperature, pump focus spot size, etc.) to the original
design, aimed at optimising the output power were made and are
elaborated in Quirrenbach et al. (1997).

\subsection{Output Beam}

In order to provide efficient excitation of the sodium in the
mesosphere, the laser must have a single spatial and temporal mode
quality.
The former criterion is achieved by ensuring that the pump lases only
in the Gaussian TEM$_{\rm 00}$ mode.
The latter is attained with a pair of low-finesse etalons and a
birefringent filter, which restrict the bandwidth to about 10\,MHz;
it can also be tuned using the scanning Brewster plate and a piezo-mounted
mirror (M3 in Figure~\ref{fig:laserbench}).
Although in principle it is possible to reduce the bandwidth to 
$\sim$1\,MHz, this is unnecessary since the 
radiative lifetime due to spontaneous emission from the 3P level of the 
atom is 16\,ns, giving a natural (homogenous) linewidth of almost 
exactly 10\,MHz.
The tuning is stabilised using a reference cavity, to which
1.4\% of the beam power is diverted.
This compares the directly-measured flux in the beam to that passing
through a temperature- and pressure-controlled etalon.
It is designed so that the 589.2\,nm line lies at the half-maximum
transmission of one of the etalon orders, providing a monotonic
relation between beam frequency and transmitted power.
For absolute frequency tuning and long-term stabilisation the cavity is
locked to the Lamb dip in the fluorescence signal from a sodium cell.
Thus the laser can be de-tuned away from the Na line for
diagnostic purposes.

\begin{figure}
\epsscale{0.9}
\plotone{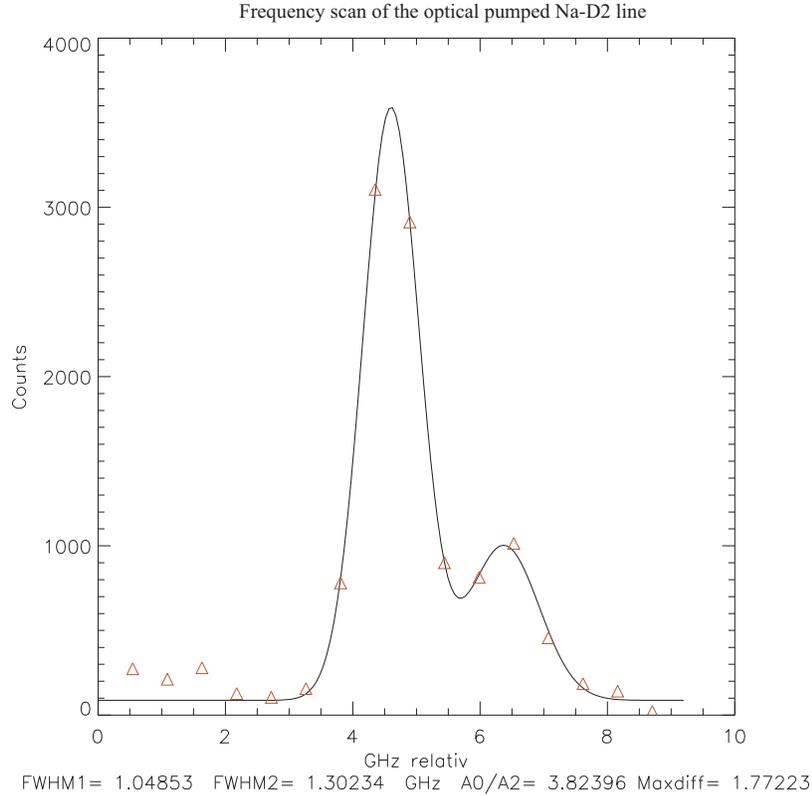}
\caption{Returned photons (LGS brightness) as a function of frequency,
during a scan across the Na D2 line. The fitted curves (representing
the absorption of the two transitions in the doublet) have a ratio of
3.8 rather than the expected 1.7 --- indicating that the laser is very
effective at pumping the lower energy transition.}
\label{fig:freqscan}
\end{figure}

Scanning the laser across the frequencies over which sodium
fluorescence is expected to occur shows an intriguing result.
It is well known that the 589.2\,nm line is a doublet with separation
1.77\,GHz, and that the lower energy transition has a 
doppler-broadened absorption cross-section greater by a factor of 1.7.
The frequency scan in Figure~\ref{fig:freqscan} reflects this, but
with an absorption cross-section ratio of nearly a factor 4;
the laser is very effective at pumping this transition.

The next component on the optical bench is a shutter (lower right in
Figure~\ref{fig:laserbench}) which
allows the laser to be shut off in an emergency, when an aircraft is
detected or when moving the telescope or dome. 

The output from the dye laser is plane-polarised, and at this stage a
quarter-wave plate circularly polarises it.
This is an important issue because, as mentioned above,
the $^2$S$_{1/2}$ ground state of the Na atom is split into a doublet
with a separation much greater than the laser line-width.
The laser is tuned to excite atoms from the higher $F=2$ angular
momentum level (with peak absorption into 3P~$F=3$).
Since they can decay back to either the $F=1$ or $F=2$ levels, after a
few cycles the whole population will reside in the $F=1$ level,
resulting in a significant loss of pump efficiency even though
the atoms are replenished by high altitude winds.
On the other hand, if the laser beam is circularly polarised so that it
imparts angular momentum to the atoms, they will tend to decay back
to the $F=2$ level, increasing the efficiency.
Measurements by \cite{ge97,ge98} at the MMT
have shown that
the brightness of a compact LGS can be increased by 30\% in this way.
There are two restrictions on the gains we can achieve.
Firstly, between the coud\'e lab and the launch telescope
there are many reflections so that the final polarisation state of
the beam is uncertain, probably elliptical.
Secondly, polarisation only becomes an important issue if
the beam is close to saturating the Na layer, otherwise the loss of
atoms from the $F=2$ state is a minor effect.
Nevertheless, the first results, given in Section~\ref{sec:pol} are
very convincing.

In the last part of the laser optical bench the beam is pre-expanded to
15--30\,mm diameter, and it is then fed into the coud\'e train.

\section{Beam Diagnostics}
\label{sec:diag}

In traversing the beam relay from the coud\'e lab to
the launch telescope, the beam suffers power loss and wavefront
aberrations from multiple reflections and dome turbulence. 
It is therefore essential to monitor the quality of the projected beam
immediately prior to its launch.
For this reason, a diagnostics bench has been installed directly
beneath the launch telescope.

\begin{figure}[h]
\centerline{\epsfig{file=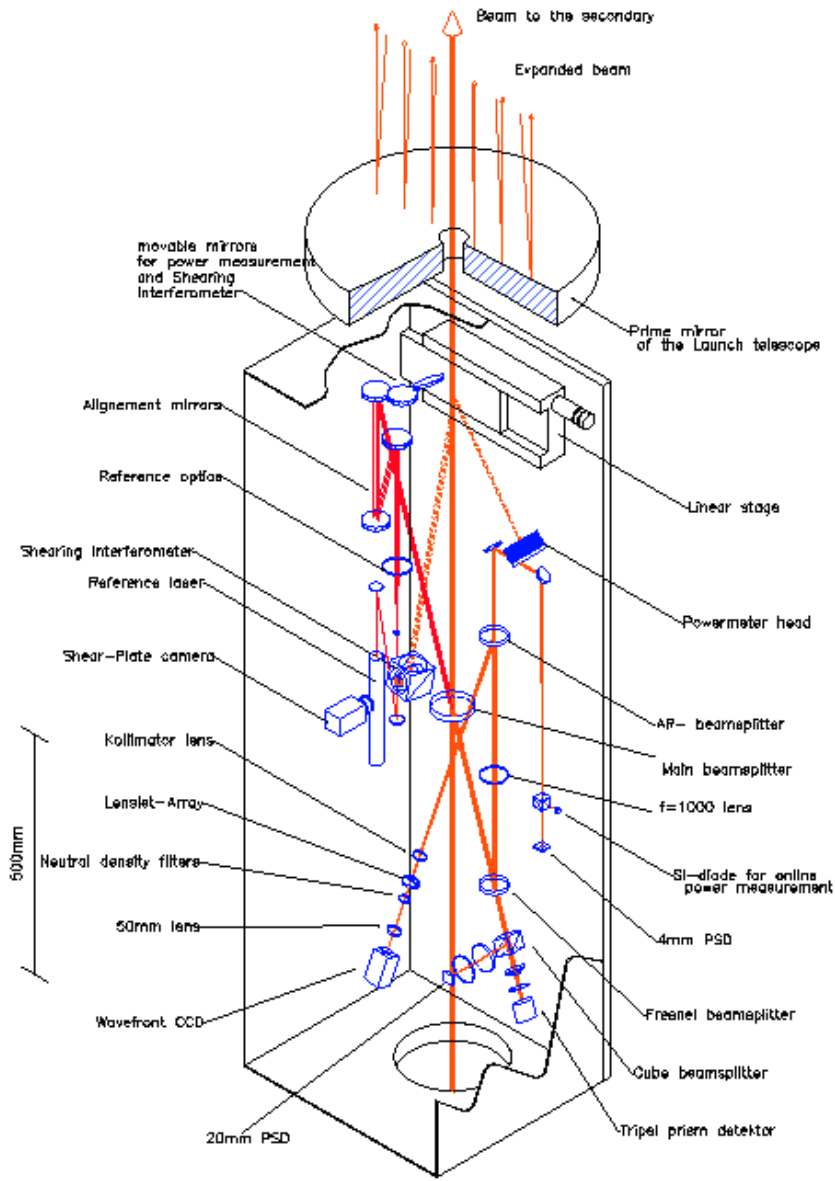,width=11cm}}
\caption{Schematic diagram showing the optical layout of the
diagnostics bench mounted directly below the launch telescope.}
\label{fig:diagnostics}
\end{figure}

Before describing the individual components, it is worth mentioning the
general layout of the system which is shown in 
Figure~\ref{fig:diagnostics}.
For the different tasks the instruments have to be fed with light from
the laser beam, but without disturbing the propagation to the launch
telescope and without noticeably reducing the power. 
For this purpose the analysis is performed on a small amount of light
which is split off with a high quality, antireflection coated optical
wedge plate at the center of the optical bench; and the polarization
detector, which is mounted at the top of the beam expander, is built as
small as possible. 
The diverted beam feeds all instruments which are continuously
needed for beam control and online measurements. 
Several mirrors are mounted on a linear stage and can be moved into the
laser beam, to do measurements that are only needed from time to time. 
Because the diameter of the laser can be adapted with a pre-expander in
the coud\'e lab to different seeing conditions, the instruments have
been designed so that they are not influenced by these changes. 

\subsection{Position measurement}

\begin{figure}[h]
\epsscale{1}
\plotone{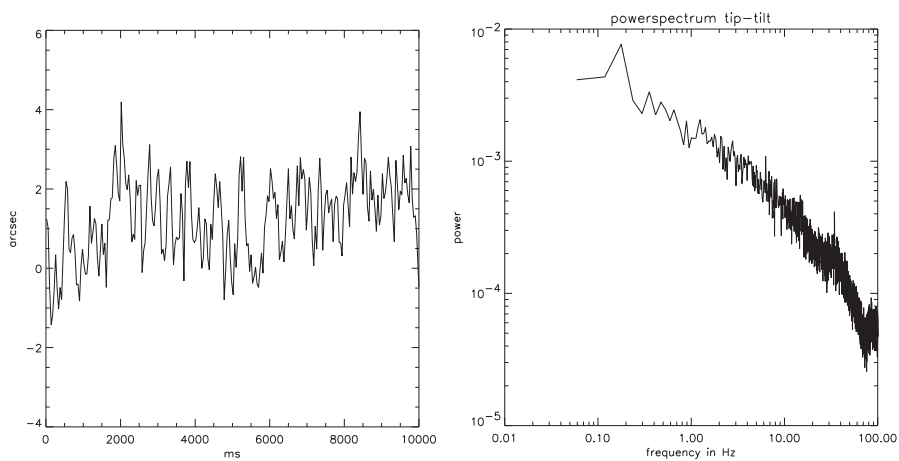}
\caption{Left: Jitter in the laser beam measured at the diagnostics
bench over a period of 10\,s, with the beam-relay control loop open.
Right: power spectrum showing that the worst jitter 
occurs with frequencies less than a few Hertz.
Note that after expanding the beam through the launch telescope, the
angular deviations are reduced by the expansion factor. 
}
\label{fig:jitter}
\end{figure}

Operation of the laser guide star requires the laser to remain steady
on the optical axis of the launch telescope, but several different
influences can affect this:
\begin{enumerate}
\item
Errors can occur when feeding the beam into the coud\'e path.
\item
The coud\'e path changes while the telescope is tracking.
\item
Turbulence in the dome may cause fast deviations in the beam position.
\item
There will be some flexure in the telescope structure and instabilities
of mechanical components.
\end{enumerate}
To minimize beam degradation due to these effects, active control of
the beam position is needed. 
As discussed in Ott et al. (this issue), the beam path is equipped
with four fast steering mirrors to keep the laser on the optical axis. 
As well as these, to position the laser accurately on the optical axis
of the launch telescope, detectors for all four degrees of freedom are
installed at the analysis bench. 
One two dimensional PSD at the focus of a 1\,m focal length lens is
capable of measuring angular deviations down to 0.4\arcsec. 
The x-y position is measured with a triple prism splitter, positioned
out of the focal plane of a short focal length lens, acting rather like
a quad-cell. 
In addition a further PSD at a 100\,mm focus is used for coarse
alignment, in case the laser is moved out of the field of view of
the other much more sensitive detectors. 
To provide an absolute reference, a helium-neon laser is aligned to the
optical axis of the launch telescope, so the detectors can be
calibrated with the signal from this light.
Measurements of the laser movement and tip-tilt measurements of the
back scattering from the mesospheric sodium can be compared with the
help of these tools, and a distinction between the influences of
atmospheric turbulence and the turbulence in the telescope dome is
possible. 
An example of the tip-tilt movement and the power spectrum (with the
beam  relay control loop open) is given in Figure~\ref{fig:jitter}.

\subsection{Power}

From the laser laboratory to the launch telescope 6 mirrors and 3
windows are installed in the telescope dome. 
Due to dust the mirrors surface quality tends to degrade with time, so
the overall transmission of the relay system decreases. 
To have a monitor for this and to know the absolute power of the light
propagating upward, a bolometric power meter is installed, to which the
laser can be fed with the help of a dielectric, spherical mirror on the
linear stage. 
Because this power measurement is not in use permanently an additional
photo diode serves as a relative online power monitor. 
The typical power of the laser at the launch telescope is about 2.5\,W,
according to an overall transmission of the relay path of 60\%. 
But this number may vary from 50\% with strong dust coverage, to 75\%
with freshly cleaned mirrors.

\subsection{Shearing interferometer}
  
\begin{figure}
\plotone{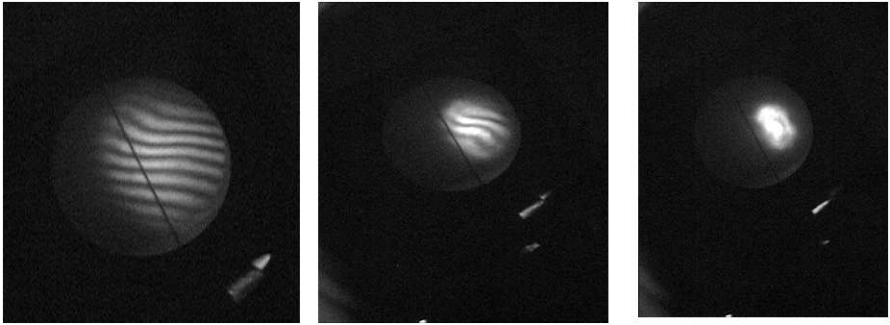}
\caption{Fringes on the Shearing interferometer at 3 stages (left
to right) during beam collimation.
When the fringes are aligned parallel to the dark line bisecting the
interferogram, collimation is perfect.
Additionally, the dominant high-order aberrations can be recognised
easily from the shape of the fringes.}
\label{fig:shear}
\end{figure}

The diameter of the laser is adapted to the actual seeing conditions
with a beam expander at the laser lab, requiring the beam to be
re-collimated after every adjustment. 
In order to do this quickly and easily, another mirror at the linear
stage can reflect the light to a shearing interferometer which is
watched by a camera.
The angle of the fringes indicates how well collimated the beam is,
with perfect collimation occuring when they lie parallel to the dark
line bisecting the interferogram, as in Figure~\ref{fig:shear}.
In addition, the shape of the fringes indicates whether abnormal
{\it static} distortions are present;
in this case they suggest significant aberrations arising from optical
elements in the beam path, such as the window at the end of the pipe
from the coud\'e lab.

\subsection{Wavefront sensing and beam profile}

\begin{figure}
\plotone{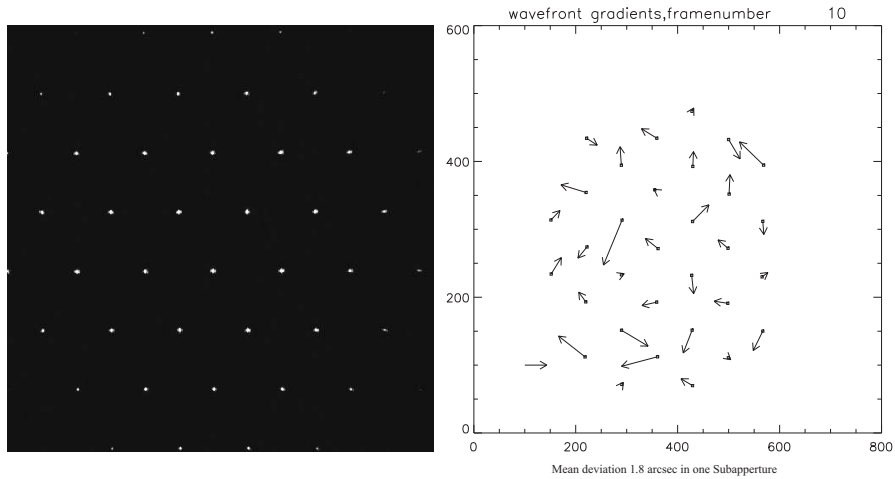}
\caption{An example of the wavefront gradients in the up-going laser,
measured with a Shack-Hartmann sensor.
Left: the pattern of spots on the WFS; 
Right: residual gradients after tip-tilt has been removed.}
\label{fig:wfs}
\end{figure}

For the higher orders of the wavefront distortion we have installed a
Shack-Hartmann sensor at one analysis beam. 
The sensor consists of a lenslet array in a collimated beam of 3--5\,mm
with arrays of $3\times3$ to $11\times11$ subapertures, one relay
lens, one neutral density filter and a CCD camera in the focal plane. 
With this sensor, direct measurements of the {\it dynamic} wavefront
distortions can be made before the laser is finally launched (static
aberrations cannot  be measured due to lack of a suitable reference beam).
Preliminary measurements, such as those in Figure~\ref{fig:wfs}, showed
an unexpectedly large distortion which reached mean values of up to
0.7$\lambda$ (at 589\,nm).
However, more recent measurements indicate that these are untypical
and that the normal wavefront error is only 0.1$\lambda$,
far too small to account for the observed LGS size.

The intensity distribution across the beam diameter, which is also an
interesting parameter in propagation calculations, can be measured with
the same CCD camera with better spatial resolution by removing the
lenslet array.
An additional CCD camera is planned, so that the intensity profile and
the wavefront can be measured simultaneously.

\subsection{Polarization}
\label{sec:pol}

\begin{figure}
\plotone{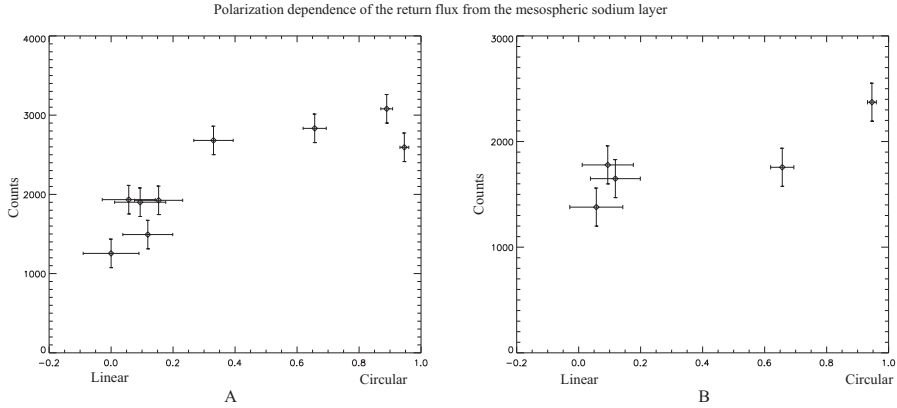}
\caption{Counts measured in the LGS as a function of polarisation, from
linear on the left of each figure, to circular on the right.
The two figures show the same experiment performed on different
occasions.
The increase in returned flux when the launch beam is circularly
polarised is extremely clear.}
\label{fig:pol}
\end{figure}

The polarization dependence of the resonant backscattering in sodium
vapor and the mesospheric sodium layer has been reported by several
authors (eg. 
\opencite{sho90};
\opencite{bal80};
\opencite{hap72};
\opencite{haw55}). 
For studies on these effects and for the automatic angular control of a
quarter wave plate a photo polarimeter has been constructed and
installed. 
Two types of instruments have been tested: 
A division of amplitude polarimeter, and one with a rotating analyzer. 
Both instruments are capable of determining a set of parameters that
describe the complete state of polarization of any elliptically
polarized light. 
While the advantage of the rotating analyzer lies in easier
calibration, the less complicated mechanics and lower sensitivity to
intensity fluctuations made the division of amplitude polarimeter the
choice to install at the exit of the launch telescope.
First measurements of the dependence of the sodium response to the
polarization were made in September 1998, and are presented in
Figure~\ref{fig:pol}.
They show the remarkable result that as much as a factor 2 increase in
returned flux can be achieved by circularly rather than linearly
polarising the projected beam.

\section{Planned Upgrades}
\label{sec:upg}

There are numerous problems associated with using mirrors to direct
the laser beam to the telescope: alignment, active mirror
control, jitter, turbulence, power loss from multiple reflections.
The most limiting remaining difficulty is wavefront curvature in the
projected beam, which is almost entirely due to distortion from the
primary mirror of the launch telescope.
A new telescope is being built for this purpose, which has been
designed to allow easy alignment while still in situ on the main
telescope.

As an alternative to the beam relay we are experimenting with using a
fibre for the same purpose.
Although this avoids many of the difficulties we have encountered, it
presents a 
new set mainly concerned with transmitting high powers through a narrow
fibre without compromising beam quality.
Experiments so far have been with a fused silica fibre which does not
preserve the polarisation, but it is possible to buy fibres which do, or
to reset the polarisation at the launch telescope.
In order to retain the mono-mode quality of the beam it is necessary to
use fibre with a core diameter of only 4\um, but this has knock-on
effects such as increasing the Brillouin backscatter to as much as
50\%; we have measured a backscatter of 1.7\,W with an input power
of 3.5\,W.
It arises from the build-up of standing acoustic waves
when high powers are forced into a narrow fibre, but
can be suppressed by phase modulating the input beam at a
frequency of about 100\,MHz.
We have successfully transported the laser beam through the fibre with
total inward and outward losses reducing the coupling efficiency to
about 60\%, similar to that of our current beam relay.
However, the energy intensity, on the order of 10$^{11}$\,W\,m$^{-2}$,
is so great that heating warps the fibre and the coupling is lost after
a few minutes.

In a similar vein we are considering installing a
Raman-fibre laser, which may become available with the required
characteristics in the next few years.
Calculations show that in order to be competitive with our current
laser and produce a \mv=9\,mag guide star, a launch power of at
least 5\,W is necessary.
The difference is due to the broader line-width ($\sim$2\,GHz) of the fibre
laser, so that much of the power is at wavelengths with a much smaller
sodium absorption cross-section (the doppler broadened profile of Na at
$\sim$215\,K in the mesosphere is 1.1\,GHz). 
In principle, such a broad bandwidth could excite both D lines which
have a separation of only 1.77\,GHz.
The maximum excitation occurs if the peak laser power is shifted by
280\,MHz from the 3S~F=2 level, although the gain would be less than
5\% since much of the power is then at frequencies where
the excitation cross-section is relatively small.
Perhaps one of the most important advantages is that much higher powers
can be used before saturation occurs, and a launch power of 25\,W could
produce a 1$''$ diameter $m_{\rm V}=7.2$\,mag LGS with almost no
saturation loss.
In practice, the difficulty is in producing such high laser powers.

\section{Conclusion}
\label{sec:conc}

The laser and beam-relay for the ALFA laser facility are nearing
completion, although they are already functioning well enough for
science observations to be carried out.

The laser beam diagnostics is opening a wide field of interesting
experiments on laser propagation in the atmosphere and effects of the
sodium response in the mesosphere. 
For the operation of the laser guide star it serves as a tool to make
the AO observations with the laser more efficient. 
Since the installation of the components in summer 1998, first
measurements with the system have started.  
Ongoing experiments will show what innovations are needed for the most
effective creation of the laser guide star for the adaptive optics.





%
%

\begin{acknowledgements}

The authors extend many thanks to the Calar Alto staff for their help
and understanding during the construction of the ALFA laser.
RID acknowledges the support of the TMR (Training and Mobility of
Researchers) programme as part of the European Network for Laser Guide
Stars on 8-m Class Telescopes.

\end{acknowledgements}

%
%

%

%

\bibliographystyle{klunamed}
\bibliography{/afs/mpa/home/davies/reference}


\end{article}

\end{document}